# A WISPR of the Venus Surface: Analysis of the Venus Nightside Thermal Emission at Optical Wavelengths

J. Lustig-Yaeger[1], N. R. Izenberg[1], M. S. Gilmore[2], L. C. Mayorga[1], E. M. May[1], A. Vourlidas[1], P. Hess[3], B. E. Wood[3], R. A. Howard[1], N. E. Raouafi[1], and G. N. Arney[4]
[1] The Johns Hopkins University Applied Physics Laboratory, 11100 Johns Hopkins Road, Laurel, MD 20723, USA; jacob.lustig-yaeger@jhuapl.edu
[2] Department of Earth and Environmental Sciences, Wesleyan University, 265 Church Street, Middletown, CT 06459, USA
[3] Naval Research Laboratory, Space Science Division, Washington, DC, USA
[4] NASA Goddard Space Flight Center, Greenbelt, MD, USA

## Abstract

Parker Solar Probe (PSP) conducted several flybys of Venus while using Venus' gravity for orbital adjustments to enable its daring passes of the Sun. During these flybys, PSP turned to image the nightside of Venus using the Wide-field Imager for Solar PRobe (WISPR) optical telescopes, which unexpectedly observed Venus' surface through its thick and cloudy atmosphere in a theorized, but until-then unobserved near-visible spectral window below 0.8 $\mu$m. We use observations taken during PSP's fourth Venus gravity assist flyby to examine the origin of the Venus nightside flux and confirm the presence of this new atmospheric window through which to observe the surface geology of Venus. The WISPR images are well explained by emission from the hot Venus surface escaping through a new atmospheric window in the optical with an overlying emission component from the atmosphere at the limb that is consistent with $O_2$ nightglow. The surface thermal emission correlates strongly with surface elevation (via temperature) and emission angle. Tessera and plains units have distinct WISPR brightness values. Controlling for elevation, Ovda Regio tessera is brighter than Thetis Regio; likewise, the volcanic plains of Sogolon Planitia are brighter than the surrounding regional plains units. WISPR brightness at 0.8 $\mu$m is predicted to be positively correlated to FeO content in minerals; thus, the brighter units may have a different starting composition, be less weathered, or have larger particle sizes.

*Unified Astronomy Thesaurus concepts:* Venus (1763); Planetary science (1255); Planetary geology (2288); Planetary surfaces (2113); Surface composition (2115); Solar system astronomy (1529); Solar optical telescopes (1514); Planetary atmospheres (1244); Solar instruments (1499)

*Supporting material:* figure set

## 1. Introduction

The Wide-field Imager for Solar PRobe (WISPR; Vourlidas et al. 2016) is the sole imager on board the Parker Solar Probe (PSP; Fox et al. 2016; Raouafi et al. 2023) mission, which was designed to study the solar corona. PSP resides in a heliocentric orbit with an aphelion around the orbit of Venus and a perihelion designed to gradually decrease from $35\,R_\odot$ to $9.86\,R_\odot$ over the course of seven years via seven Venus gravity assists (VGAs).

As with previous interplanetary missions, VGAs provide an opportunity to conduct brief, but impactful science while at Venus. During VGA3 on 2020 July 11 and VGA4 on 2021 February 20, PSP/WISPR was used to observe the nightside of Venus, and the resulting broadband optical images revealed a surprising sensitivity to the surface. Wood et al. (2022) provided initial and novel evidence that the WISPR flyby observations of the Venus nightside are sensitive to the thermal emission from the hot Venus surface at wavelengths shortward of 0.8 $\mu$m. While sensitivity to the Venus surface at red-optical wavelengths had been previously predicted (Moroz 2002; Knicely & Herrick 2020), the PSP/WISPR observations demonstrated it clearly and present the shortest wavelength Venus nightside observations to date (Wood et al. 2022).

Venus' thick and cloudy atmosphere poses a formidable challenge to remote sensing observations of the subcloud region. However, thermal emission from the hot surface is able to emerge from the Venus atmosphere through discreet opacity windows between $CO_2$ and $H_2O$ absorption bands, where it can be observed on the planet's nightside. As a result, there is a rich scientific literature of using these nightside opacity windows to conduct studies of the Venus lower atmosphere (e.g., Allen & Crawford 1984; Crisp et al. 1989; Pollack et al. 1993; Meadows & Crisp 1996; Arney et al. 2014; Peralta et al. 2017; Iwagami et al. 2018) and surface (e.g., Hashimoto & Sugita 2003; Hashimoto et al. 2008; Mueller et al. 2008; Smrekar et al. 2010; Basilevsky et al. 2012; Gilmore et al. 2015; Shalygin et al. 2015). Over time, opacity windows have been discovered at ever shorter wavelengths where the thermal flux from Venus is a lower contrast, and the signal-to-noise requirements are higher. The nightside thermal brightness of Venus rapidly decreases toward shorter wavelengths in the near-IR (NIR) and optical, thus requiring higher precision instruments and flyby spatial resolution to detect. Notably, observations made during VGA flybys have led to critical insights into Venus nightside remote sensing, and include flybys by Galileo in 1990 (Carlson et al. 1991; Drossart et al. 1993; Hashimoto et al. 2008), Cassini in 1999 (Baines et al. 2000), and now PSP in 2020, 2021, and forthcoming in 2024.

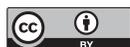

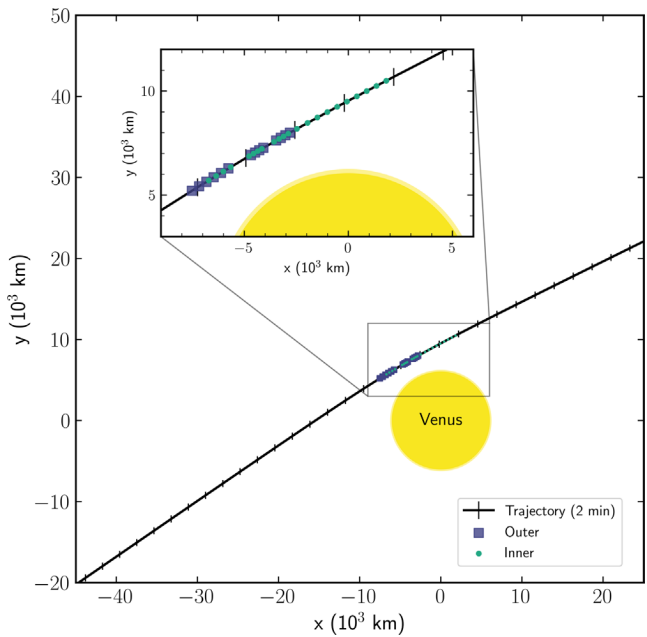

**Figure 1.** PSP's trajectory during VGA4 (2021 February 20) with the location of the spacecraft during WISPR-O and WISPR-I observations denoted with purple squares and magenta circles, respectively. Tick marks are displayed every 2 minutes along the trajectory, and a slight deflection is seen at Venus due to the change in PSP's orbit from the gravity assist. The Sun located in the $-y$-direction.

In this paper, we build on the work of Wood et al. (2022) and examine the potential for Venus surface science using the WISPR flyby observations. We correlate WISPR emission to mapped geologic units to assess the sensitivity of WISPR observations to potential surface compositional differences; simulate thermal emission from the surface through the thick and cloudy Venus atmosphere across a broad range of surface temperatures and emission angles to provide robust evidence that surface thermal emission dominates the observations; and discuss potential contributions to observed spatial variability. In Section 2, we describe our methods, including descriptions of the WISPR instrument and data, how we use World Coordinate System (WCS; Thompson & Wei 2010) and Spacecraft Planet Instrument Camera matrix Events (SPICE; Acton 1996; Acton et al. 2018) information to project various Venus data sets to the WISPR perspective, and our atmospheric radiative transfer model. In Section 3, we present our results, including the validation of the in-band photometry against any possible light leak, how the WISPR images correlate with known Venus surface for specific known geologic units, and provide a comprehensive model explanation and reproduction of the WISPR images. In Section 4, we discuss the implications of our findings for current and future explorations of Venus and discuss limitations. We conclude in Section 5.

## 2. Methods

### 2.1. Instrument and Data

WISPR consists of two telescopes, WISPR-Inner and WISPR-Outer (Vourlidas et al. 2016), hereafter referred to as WISPR-I and WISPR-O, respectively. Both telescopes use Active Pixel Sensor detectors with $2048 \times 1920$ pixels and have broad optical bandpass covering roughly 0.48–0.80 $\mu$m (Vourlidas et al. 2016; Hess et al. 2021).

In this paper, we focus exclusively on the 2021 February fourth PSP flyby of Venus. Figure 1 shows a top-down view of the PSP trajectory during VGA4 (in the J2000 SPICE reference frame). Based on the relative field of views (FOVs) of each WISPR telescope, WISPR-O began observing Venus prior to WISPR-I, and WISPR-I continued observing Venus after WISPR-O. For this work, we study only the WISPR-O images of Venus during VGA4. Although we were able to generally reproduce the major findings in this paper using WISPR-I data, errors in the projection of reference data onto the image plane (see Section 2.3) limit the accuracy of our latitudinal and longitudinal knowledge on Venus in the WISPR-I images such that it is difficult to precisely identify correlations between WISPR brightness and known surface properties (e.g., elevation). An analysis of the WISPR-I data will therefore be the subject of future work. Figure 2 shows the sequence of WISPR-O images observed during the 2021 Venus flyby (VGA4). A figure set showing analogous flyby images for WISPR-I is available in the online journal.

### 2.2. WISPR Reduction

As with Wood et al. (2022), we use the WISPR level 2 images of the flyby. These images are calibrated in units of mean solar brightness (MSB) according to the procedure detailed in Hess et al. (2021). To convert between MSB and DN s$^{-1}$, a calibration factor of $C_f = 9.24 \times 10^{-14}$ is used for WISPR-O based on the in-flight star calibration (Hess et al. 2021), but modified to account for the gain setting used for the Venus observations.

Beginning with the level 2 calibrated data products, we remove residual striping in the images along the readout direction, likely due to light smearing during the readout, since WISPR cameras lack a shutter. We first mask the pixels that contain Venus' disk, as well as 20 pixels on each side around the edges of the frame that are impacted by reflection off the protective barrier at the edge of the detector (edge effects were treated similarly for WISPR-I images in Stenborg et al. 2021). The median of the remaining pixels in each row is then calculated and subtracted from that row, significantly improving the apparent striping effect, which appears constant across an entire row. This step also effectively removed the sky background from each image, which appeared due to excess scattered sunlight that was particularly apparent in the first two WISPR-O frames, but decreased as PSP entered the Venus penumbra.

### 2.3. Coregistered Reference Data and Coordinate Transformations

We use a series of reference radar data sets from the Magellan mission (frequency = 2.4 GHz, $\lambda = 12$ cm, Ford & Pettengill 1992; Ford et al. 1993) to explore correlations with known surface conditions. These include Magellan Global Topographic Data Records, Magellan Global Emissivity Data Records; each of these mosaics are resampled to a spatial resolution of 4.6 km pixel$^{-1}$, which is the standard for Magellan data products as described in the data reference[5] (Ford & Pettengill 1992; Ford et al. 1993). We also use the Magellan Synthetic Aperture Radar (SAR) FMAP Left Look Global Mosaic with a spatial resolution of 75 m pixel$^{-1}$.

---

[5] https://astrogeology.usgs.gov/search/map/Venus/Magellan/RadarProperties/Venus_Magellan_Topography_Global_4641m_v02

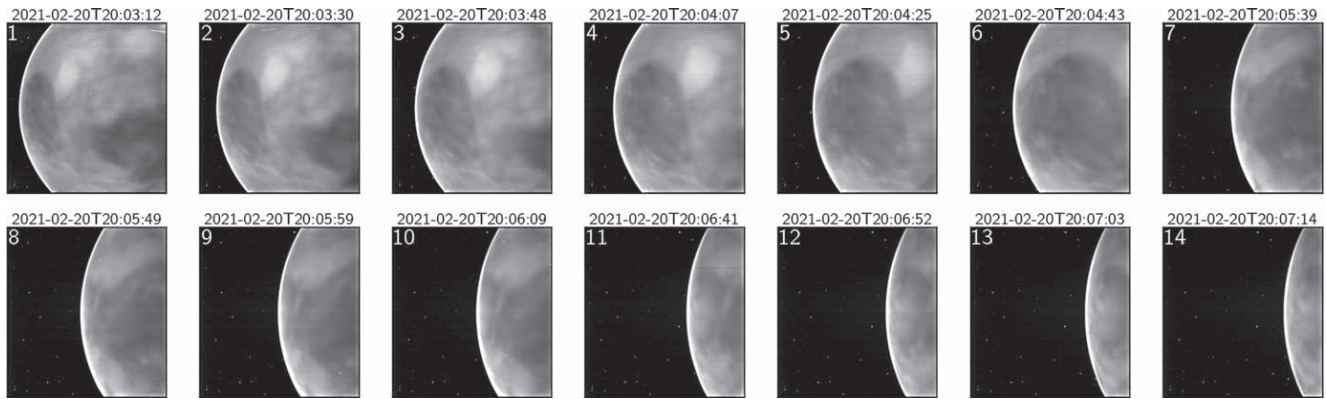

**Figure 2.** Sequence of fully reduced WISPR-O images observed during the 2021 February 20 flyby. A figure set showing analogous flyby images for WISPR-I is available in the online journal.

(The complete figure set of 2 images is available.)

Surface geological units were identified in the Magellan data on the basis of morphological characteristics in the SAR data (e.g., Brossier & Gilmore 2020). These were then mapped to exclude pixels at the edges of the units that might contain multiple types of surface materials given the spatial scale of the WISPR-O observations and motion blur. No presupposition was made of any correlation between geology and WISPR-O; instead, we aim to test for evidence of correlations between WISPR-O signatures and radar-derived quantities for surface units previously identified to be geologically distinct. Throughout this work, elevation values are quoted relative to 6051.8 km, the mean planetary radius.

The Venus reference data products are projected onto the WISPR FOV to facilitate scientific investigations. The projection follows exactly what was done by Wood et al. (2022), but now expanded to include more data products described using physical units. Since PSP is a solar orbiter, the fits files use the WCS Helioprojective-Cartesian (HPC; Thompson & Wei 2010). The date and time for each exposure is used with SPICE (Acton 1996; Acton et al. 2018) kernels to extract the HPC coordinates of each pixel. We use the Python port SpiceyPy (Annex et al. 2020). The SPICE kernels allow us to track spacecraft location and location, Venus' location, shape, and size, as well as conduct coordinate and reference frame transformation. From HPC, we convert the coordinates to Venus mean equator (VME) of date coordinates (which requires a transformation from HPC to heliocentric Aries ecliptic, and then to VME). The result is that each pixel has a resulting latitude and longitude on Venus from the computed line of sight—surface intersection point. Sky values are treated as NaNs. The maps, e.g., Venus Magellan Global Topography 4641m v2, are projected in simple cylindrical coordinates, and thus, a projected image is built up by filling in the requisite data at each latitude and longitude pixel coordinate. We use the IAU rotation period for Venus of 243.0185 days as is given in the SPICE kernels. Mueller et al. (2012) suggest that longitude offsets are on the order of $-0.3°-0.08°$, which is a smaller effect than the motion blur in the WISPR-O images.

The flyby begins with an excess of scattered sunlight in both WISPR-O and WISPR-I as PSP enters the Venus shadow. The spacecraft orientation is such that WISPR-I is essentially looking nadir at Venus' barycenter. The flyby occurs over a timescale of roughly 500 seconds, and the trajectory is shown in Figure 1. Exposures vary between roughly 3–7 s for the WISPR-I camera and roughly 4 s for the WISPR-O camera during the flyby. The spacecraft orientation remains fixed during the flyby and does not remain targeted at Venus' barycenter causing Venus to progressively *slip out* of the FOV. Thus, throughout each image, a certain amount of blurring is visible, which we had originally attributed solely to atmospheric scattering, but is also caused by the motion of the spacecraft during the duration of the exposure. Thus, we project each map at 9 different times within a single exposure time duration and average them together to reproduce this motion blur. For the maps of individual surface units, we combine all the maps to one single map where each value represents a different geologic unit. Instead of averaging the 9 subtime projections, we instead only count pixels to be in the geologic unit if they fall consistently within the unit in all subtime frames. This helps to ensure that when we evaluate by geologic unit we are only considering pixels that are not contaminated with background sky or other geologic units.

Figure 3 shows one of the WISPR-O flyby images along with the Magellan elevation and radar emissivity data projected into the frame. By visual inspection, it is clear that the WISPR-O intensity correlates (inversely) with the elevation data. The subsequent analysis of correlations between the WISPR and Magellan data sets is presented in Section 3.2.

### 2.4. Radiative Transfer Modeling

We use the Spectral Mapping Atmospheric Radiative Transfer code (SMART; Meadows & Crisp 1996; Crisp 1997) to simulate the top of atmosphere Venus radiances for comparison with the WISPR measurements given contemporary knowledge of the Venus atmosphere and surface. SMART solves the radiative transfer equation for one-dimensional plane-parallel atmospheres using line-by-line, multi-stream, multi-scattering calculations. All radiative transfer calculations in this work were made at 1 cm$^{-1}$ wavenumber resolution and used 8-streams (four upward and four downward) at Gaussian quadrature computational points, except where specified for finer sampling of observer emission angles. We used spatially averaged Venus International Reference Atmosphere (VIRA) thermal and molecular profiles from Moroz & Zasova (1997) using the lower atmosphere updates from Arney et al. (2014). Line-by-line rovibrational molecular absorption coefficients for $CO_2$, $H_2O$, CO, $H_2S$, HF, and HCl were calculated using the LBLABC code (Meadows & Crisp 1996; Crisp 1997) using the HITRAN2016 line list (Gordon et al. 2017). Sulfuric acid clouds were simulated using the nominal model from Crisp (1986) containing Mode 1, Mode 2, Mode 2', and Mode 3 particles, which are all assumed to be 75% $H_2SO_4$ by weight (Arney et al. 2014). Following the approach of Crisp (1986), we include the unknown UV absorber as a modified version of the submicron Mode 1 particles that match dayside observations of the Venus spherical albedo in the optical. All of the vertically resolved atmospheric profiles (cloud optical depths, gas volume mixing ratios, and the thermal profile) were held static at the nominal global profiles throughout this work. We explore the effect of changes in cloud opacity in Appendix, but find that large changes in cloud optical depths lead to relatively small changes in WISPR brightness. This is because the cloud model is roughly 35% more transparent in the optical than in the 1 $\mu$m window. We further discuss the implications and caveats of these atmospheric assumptions in Section 4.2.

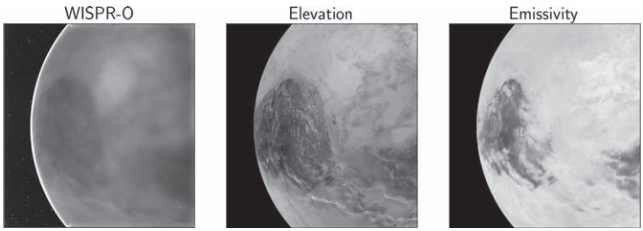

**Figure 3.** WISPR-O image (left panel) compared to Magellan elevation (middle panel) and emissivity data (right panel) projected into the WISPR-O FOV. All panels use a different relative gray-scale color scheme to emphasize visual similarities wherein the transition from black to white scales from low to high WISPR-O brightness, high to low surface elevation, and low to high surface radar emissivity. While the WISPR-O image contains numerous visual similarities with surface features seen in both reference data sets, the images have a lower spatial resolution due to intense atmospheric scattering (e.g., Moroz 2002). The date and time corresponding to this WISPR-O image is 2021 February 20 20:03:48.

Given the nominal radiative transfer modeling setup described above, we run a series of models to simulate the expected sensitivity of the WISPR observations to known and/or predictable variations in the surface conditions. We produce thermal radiances across a three-dimensional grid in (1) the elevation (temperature) of the surface, (2) the emissivity of the surface at relevant WISPR wavelengths, and (3) observer zenith angles. For the surface elevation and temperature, we use the VIRA thermal profile to define the relationship between altitude and temperature, and then, we truncated the atmosphere accordingly across a grid in elevation from −2 to 20 km at 1 km intervals. This procedure ensures that the surface temperature scales physically with systematic changes in surface elevation following the atmospheric thermal structure. While the majority of the Venus surface has elevations ⩾0 km, some of the lowest elevation regions lie below this level and therefore have negative elevation values relative to the zero-point. For the surface emissivity, we assume wavelength independent values that range from 0.5 to 1.0 at intervals of 0.05. For the observer zenith angles, we ran simulations at the finite angles of 86°.0, 70°.7, 59°.3, 21°.5, and 0°.0.

We followed the same procedure as Wood et al. (2022) to convert spectrally resolved top of atmosphere radiances to photometric counts in units of digital number per second (DN s$^{-1}$) within the WISPR bands. This conversion is given by the following convolution integral over wavelength:

$$\text{DN s}^{-1} = \frac{\Omega}{g} \int_{\lambda_0}^{\lambda_1} \frac{A_{\text{eff}} R}{E_{\text{phot}}} d\lambda \qquad (1)$$

where $A_{\text{eff}}$ is the WISPR effective area sensitivity curve for WISPR-O (or WISPR-I), $R$ is the Venus thermal emission radiance spectrum, $E_{\text{phot}}$ photon energy of each spectral interval, $\Omega$ is the angular extent of each pixel, and $g$ is the detector gain.

## 3. Results

### 3.1. Validation of Light Sources

The stark sensitivity to the Venus surface in the WISPR images initially raised concerns that it could have been explained by emission from well-known opacity windows, for example, at 1 $\mu$m being picked up by excess sensitivity beyond the nominal WISPR bandpass. Wood et al. (2022) reported on lab measurements from the spare WISPR optics that showed no signs of a red light leak. The consistency of the observed counts with thermal emission models provides a second line of evidence supporting the in-band nature of the WISPR Venus flux (Wood et al. 2022). We now present a third line of evidence to evaluate whether any light leaks might be present by looking at the background stars in each WISPR image.

Using SPICE, we query for which stars of sufficient brightness from the Hipparcos catalog (Perryman et al. 1997) should have fallen within the FOV and determined their pixel coordinates. Looking up the stars in Simbad (Wenger et al. 2000) allows us to select stars with known effective temperatures and gravities. We then compare the flux detected by WISPR with that predicted by the appropriate PHOENIX model (Husser et al. 2013) convolved through the WISPR bandpass. We show an example of a WISPR-O image and stellar spectra in Figure 4. If the WISPR bandpass has a previously uncharacterized light leak, additional flux would be measured for all stars. If the light leak were bluer than the known bandpass, only hotter bluer stars would show additional flux, because the flux from redder stars is proportionally less at these wavelengths. If the light leak were redder than the known bandpass, redder stars would be proportionally more affected than bluer stars. We conduct aperture photometry on the stars in the WISPR image and show the results of the measure versus PHOENIX model expected values in Figure 4. The lack of clear trend here is suggestive that there is no light leak. Adding a faux light leak to the WISPR transmission function used to predict the Phoenix model fluxes shows a clear trend with stellar temperature that cannot be confused with random noise. Thus, we can be confident that the observations are only sensitive to in-band photons, and a window through Venus' atmosphere to its surface is in fact present at these wavelengths.

### 3.2. WISPR Correlations with Known Surface Characteristics

Figure 5 shows the relationships between WISPR-O counts, elevation, and radar emissivity for a variety of surface geological units. Only the pixels with emission zenith angles <60° are used in this and the subsequent surface analyses unless otherwise stated to reduce the confounding effects of limb brightening, as discussed later. Five units are mapped, including the tessera terrains Ovda Regio, Thetis Regio, and

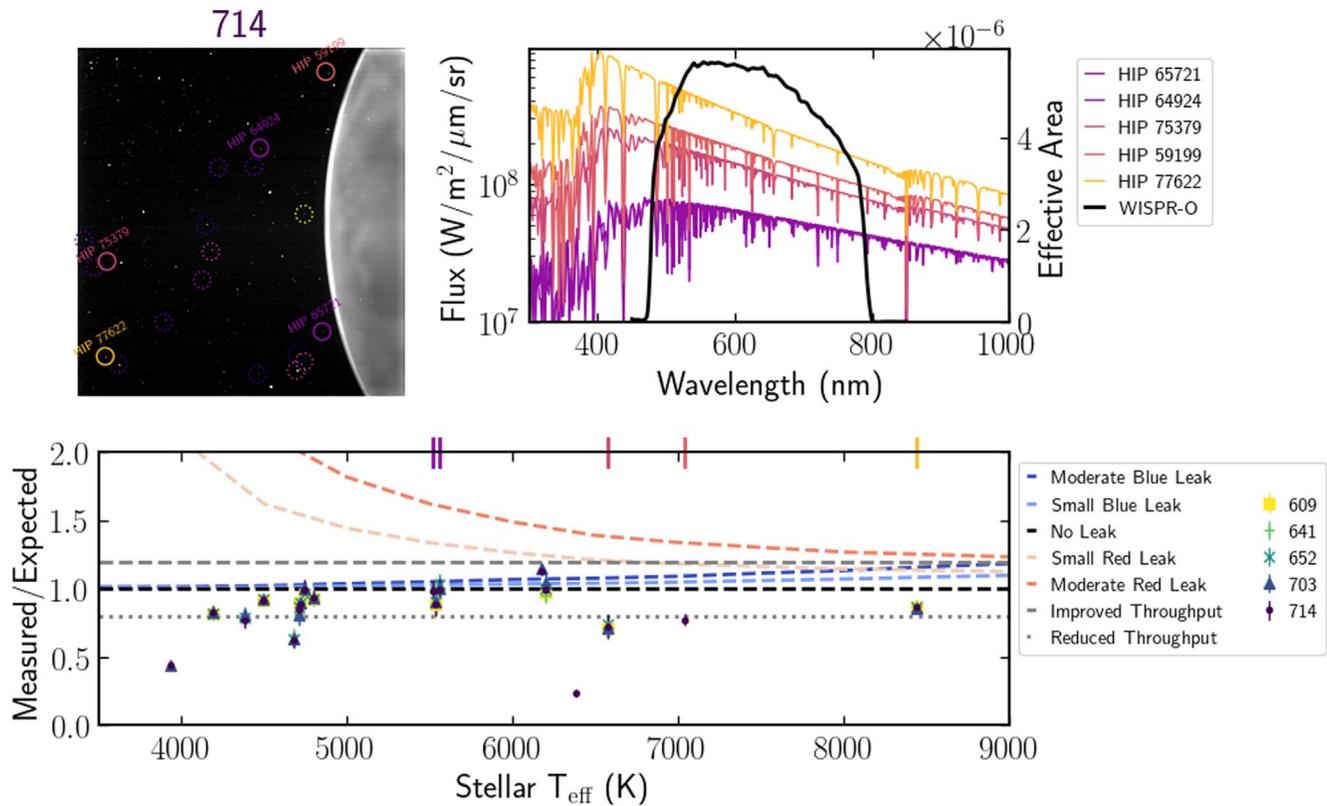

**Figure 4.** Left: WISPR-O image (200714) with pixel locations of stars that should have fallen within the field of view. Stars that are behind Venus are ultimately not considered. Right: their Phoenix model spectra as compared to the WISPR bandpass for those stars that are on the main sequence. Bottom: measured fluxes from stars in the WISPR images (denoted by color and symbol) as compared to the expected fluxes from Phoenix stellar models as a function of stellar effective temperature. Tick marks label the main-sequence stars from the upper right panel for reference. No clear pattern exists that would be indicative of a light leak in either the blue or red direction (dashed lines).

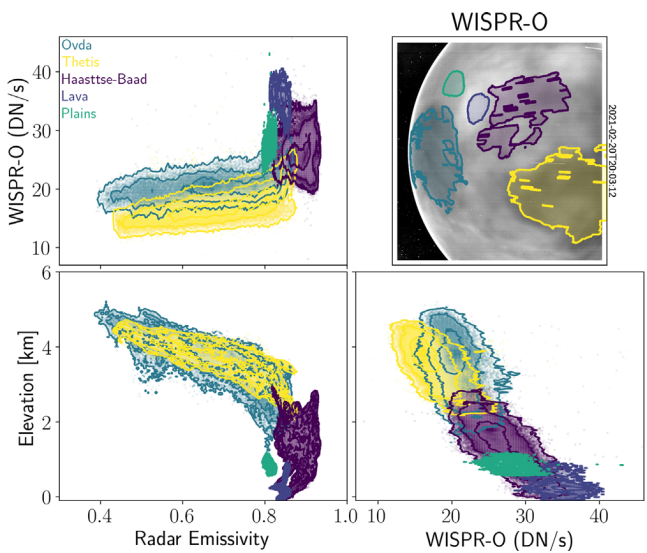

**Figure 5.** Two-dimensional contours showing correlations between WISPR-O brightness, surface elevation, and radar emissivity for a variety of different surface geological units (colors) using image data from the first frame in the 2021 flyby (upper right). Adjacent axes share the same quantities and axis limits to facilitate comparisons between panels. The lower left elevation–emissivity panel shows only reference data. Contours show the density of measurements at the $1\sigma$, $2\sigma$, and $3\sigma$ level.

Haasttse-baad tessera, and two plains units: "Lava," which has a high WISPR-O brightness signature and "Plains," which has WISPR-O values typical of regional plains. "Lava" corresponds roughly with the undivided smooth flow and shield terrains Sogolon Planitia of the Niobe Planetia quadrangle of Venus (Hansen 2009), whereas the "Plains" to the northeast are Niobe Planitia proper. The image and analysis shown is for the first frame in the 2021 flyby. Multiple interesting correlations are evident. Together, the combination of mapped geological units span a large range in altitude from about 0 to 5 km and exhibit a strong negative correlation with WISPR brightness. This is a well-known feature of thermal emission from the hot Venus surface (e.g., Mueller et al. 2008) and is a result of the Venus temperature profile, and its characteristic decrease with altitude (Seiff 1987; Lorenz et al. 2018). Ovda and Thetis are the highest elevation regions in view and correspondingly have the lowest WISPR brightness, whereas the region labeled "Lava" is the lowest elevation region in view and has the highest WISPR brightness. Although also tessera terrain, Haasttse-baad has a much lower maximum elevation than either Ovda or Thetis, and correspondingly higher WISPR-O values. The WISPR-O elevation trends of Haasttse-baad and Thetis appear colinear, and divergent from the Ovda trend.

Radar emissivity appears positively correlated with WISPR brightness; however, the relationship between elevation and radar emissivity (e.g., Figure 5 lower left) exhibits a characteristic negative correlation known for some regions of the Venus surface (e.g., Klose et al. 1992) that can propagate the surface elevation (temperature) trend into radar emissivity space, complicating the assessment of trends with radar emissivity. The negative trend of radar emissivity with elevation is ascribed to the volume and type of high dielectric minerals in surface rocks (Klose et al. 1992), while the VNIR emissivity is a function of FeO content (Dyar et al. 2020); it has not been yet shown that there is a systematic relationship between these two characteristics.

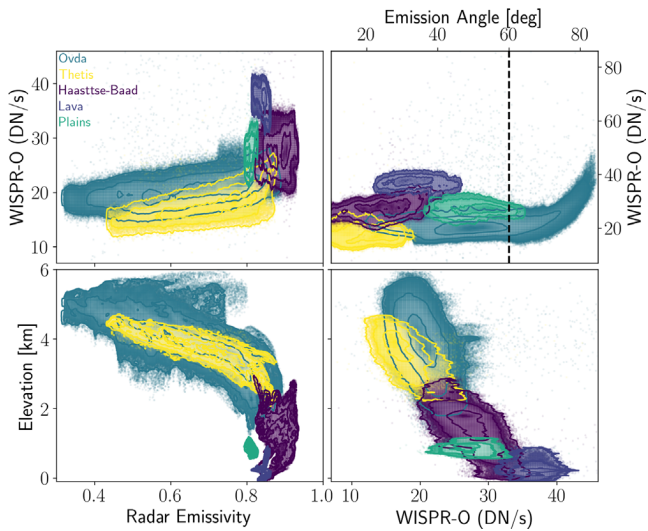

**Figure 6.** Two-dimensional contours showing correlations between WISPR-O brightness, surface elevation, and radar emissivity for a variety of different surface geological units (colors) using image data from all frames in the 2021 flyby. The upper right panel shows WISPR-O brightness as a function of emission zenith angle over a larger range in WISPR counts than shown in the other panels. The black vertical dashed line shows the maximum emission angle cut used in the other panels to reduce the confounding limb brightening effects from the assessment of surface information. Contours show the density of measurements at the $1\sigma$, $2\sigma$, and $3\sigma$ level.

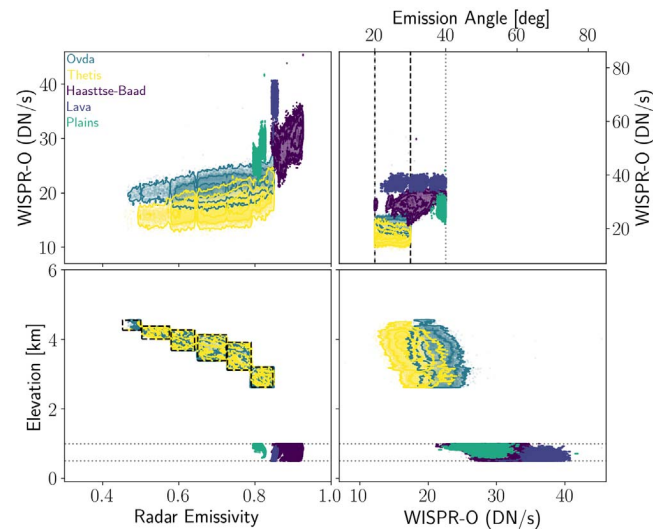

**Figure 7.** Similar to Figure 6 except focusing on data with specific emission angles, elevations, and emissivities to facilitate comparison and reduce confounding factors. The two high elevation geological units, Ovda and Thetis, are shown only for data in the series of elevation–emissivity boxes where the two units have significant overlap, and for emission angles between 20° and 30° where the two overlap. The low elevation units, Haasttse-baad, Lava, and Plains, are shown in a narrow range of elevations between 0.5 and 1 km and for emission angles between 20° and 40° where they overlap. As before, contours show the density of measurements at the $1\sigma$, $2\sigma$, and $3\sigma$ level.

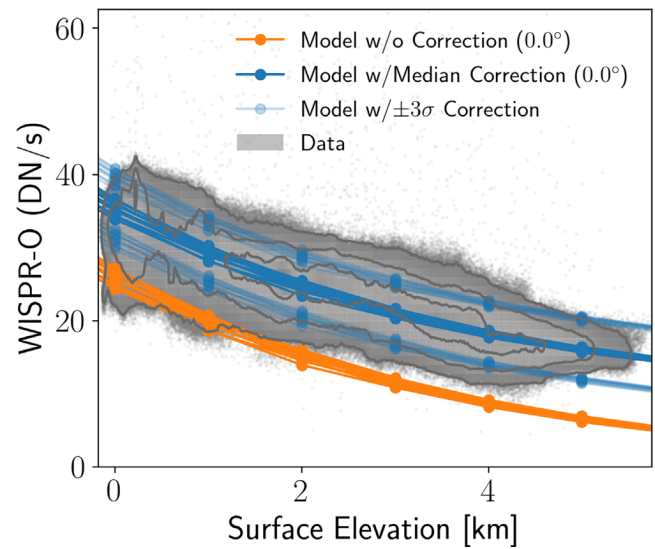

**Figure 8.** WISPR-O count rates as a function of surface elevation (gray points and contours). Our baseline nadir thermal emission models for a variety of thermal emissivity cases are shown in orange. Thermal models corrected by an empirical brightness offset (9.5 DN s$^{-1}$) are shown for the median value (blue) and the $\pm 3\sigma$ range (semitransparent blue). Contours show the density of measurements at the $1\sigma$, $2\sigma$, and $3\sigma$ level. The baseline thermal models underpredict the WISPR-O brightness, and the empirical correction places them in excellent agreement.

Similar to Figure 5, Figure 6 shows the relationships between WISPR-O counts, elevation, and radar emissivity for the same geological units, but now shown using 11 frames from the 2021 Venus flyby. Many of the same features of Figure 5 are seen in Figure 6 although now supported by a significantly larger quantity of measurements. In Figure 6, the relationships between the tessera units are even more distinct, including the Haasttse-baad–Thetis colinearity trend in WISPR-O versus elevation, and the divergence of Ovda from that trend. The distinction between the Lava unit and the Plains is also emphasized in this expanded data set. The relationship between WISPR-O counts and the emission zenith angle is also shown in the upper right of Figure 6. In general, WISPR brightness for a given terrain remains relatively flat at emission angles up to 60°–70° and then rises sharply toward the limb at 90°. This limb brightening trend motivates our choice to restrict the surface analyses to emission zenith angles of 20°–40°.

Figure 7 is a restricted subset of Figure 6, highlighting very narrow regions of the parameter space to facilitate comparison between geological units. We compare units at the same elevation to remove the temperature effect on emissivity and limit WISPR-O emission angles to 20°–40° to consider surface emission viewed with a similar path geometry through the atmosphere. When Ovda and Thetis regio are compared this way, Ovda has a brighter WISPR-O signature than that of Thetis by ∼20% at 1.6σ (or 3.6 DN s$^{-1}$ on average) where the two tesserae have both similar elevation and radar emissivity. The Lava unit is brighter than the Plains unit by ∼35% at 4σ (or 9.6 DN s$^{-1}$ on average) at similar elevations and with only small differences in radar emissivity (∼0.05).

### 3.3. Comparison with Surface Thermal Emission Models

Figure 8 shows the WISPR-O brightness as a function of surface elevation compared to our baseline nadir thermal emission models (orange lines) over the same range of elevations. WISPR-O data are used from all VGA4 flyby images, but only those with emission zenith angles <40°. The WISPR-O counts are visibly negatively correlated with elevation and yield a correlation coefficient of −0.81 with a $p$-value of zero. Although the models exhibit the same trend with elevation, they are clearly offset relative to the measurements by an apparent constant offset, indicating either an additional source of light or an overestimation of the atmospheric opacity in the models. While thinner clouds than those from the nominal model used in our radiative transfer calculations could cause the thermal flux from the surface to be enhanced, as shown in Appendix, the effect is minimal for modest changes in cloud opacity. We note that this could also be a product of imperfect calibration factors used to convert between physical flux units and observed counts. However, since the WISPR images in Figure 2 clearly show a bright limb and limb brightening is apparent at high emission angles in Figure 6, likely due to $O_2$ and $O$ I nightglow (Wood et al. 2022), we entertain the hypothesis that there is additional nightglow across the entire Venus disk.

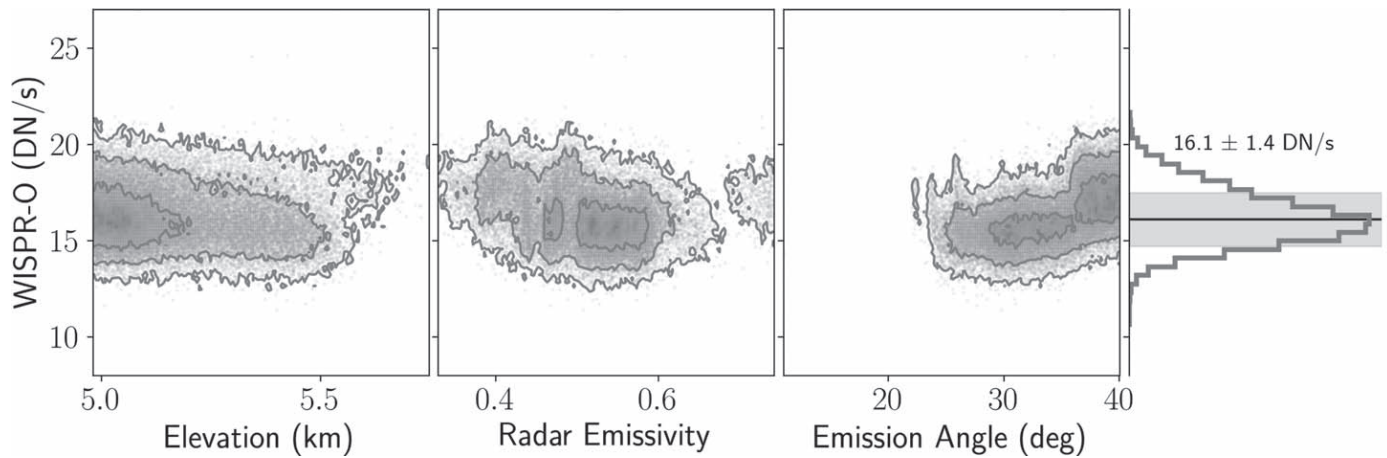

**Figure 9.** WISPR-O counts for pixels that are expected to be low brightness based on high elevation (low surface temperature) and low emission angles (no limb brightening). Counts are shown as a function of elevation (first panel), radar emissivity (second panel), and emission angle (third panel). Contours show the density of points at the $1\sigma$, $2\sigma$, and $3\sigma$ level. The rightmost panel shows a histogram of WISPR-O counts for the set of points, which show an average brightness of $16.1 \pm 1.4$ DN s$^{-1}$, nearly 9 DN s$^{-1}$ in excess of the expected brightness from model predictions.

The pixels with an expected low brightness from the thermal emission models provide a means to measure the average nightglow brightness at relatively low emission angles. Figure 9 shows the WISPR-O counts for all high elevation (>5 km), low emission angle (<40°) points from all VGA4 flyby images as a function of elevation, radar emissivity, and emission angle, and compressed into a one-dimensional histogram. This subset of points with expected low intensity surface thermal emission have a WISPR-O count rate of $16.1 \pm 1.4$ DN s$^{-1}$. Since our thermal models still predict a measurable surface emission for regions at and above 5 km, we subtract off the 5 km predicted flux (6.6 DN s$^{-1}$) from the mean brightness excess derived in Figure 9 to obtain an offset of $9.5 \pm 1.4$ DN s$^{-1}$. We take this value to be our empirical estimation for the excess brightness near the Venus disk center, which may ultimately be due to a combination of nightglow emission, thinner clouds than our nominal model, scattered sunlight from the dayside, or a flux calibration offset.

Using the empirical offset to correct missing radiative processes in our thermal emission spectral models can help to explain the brightness levels observed with WISPR. Figure 8 shows nadir thermal emission models convolved with the WISPR-O bandpass with the empirical brightness correction added (blue and teal lines). The overlapping lines indicate a range caused by thermal emissivity variations between the models, whereas the faint blue models show the $\pm 3\sigma$ bounds around the median empirically corrected models. The fit to the elevation trend is significantly improved using the empirical offset. Therefore, the empirical brightness correction derived using only high elevation pixels helps to explain both the offset and breadth in the scatter of the measured trend with elevation relative to our thermal calculations. Furthermore, the consistency between the thermal models and the WISPR observations robustly validates the nature of the observations as thermal emission from the surface.

### 3.4. Modeling the WISPR-O Images

Extending beyond the insights gleaned from our analysis of empirical trends in the WISPR-O data, we use our precomputed grid of radiative transfer simulations of the Venus thermal emission to construct a full model of the WISPR-O images. We linearly interpolate the thermal radiance models onto the WISPR image projections, taking two-dimensions into account. The first is the WISPR-projected Magellan altitude data to capture the temperature-dependence of the radiance, and the second is the map of subspacecraft emission zenith angles to capture the angle dependence of atmospheric path lengths for rays emitted from the surface. We hold the thermal emissivity of the surface fixed at 0.9 consistent with a surface albedo of 0.1 measured by Venera 9 and Venera 10 (Ekonomov et al. 1980).

Figure 10 shows the third WISPR-O image from the 2021 flyby on the leftmost panel compared to our thermal emission only WISPR-O image model in the second panel. The thermal model captures much of the spatial variability in brightness contrast, which is consistent with the previously determined strong correlation between the altitude (temperature) of the surface and the WISPR brightness. However, the thermal model exhibits limb darkening rather than the stark limb brightening seen in the true image. These characteristics are consistent with the lack of nightglow emission in the spectral models.

A small linear correction was applied to our simulated surface thermal emission image to optimally account for the unknowns associated with cloud opacity and a baseline flux offset. We determine a scale factor (slope) and baseline offset (intercept) that when applied to our nominal thermal emission models minimizes the residuals with the WISPR-O image for pixels with emission angles $<35°$ of the disk center. This yielded a thermal flux scale factor 0.95 and an offset 11.18 DN s$^{-1}$. This offset is similar to the empirically derived flux offset in Section 3.3. The thermal model panel in Figure 10 incorporates the multiplicative factor, while the the additive offset is visualized in the third panel along with the nightglow component discussed next.

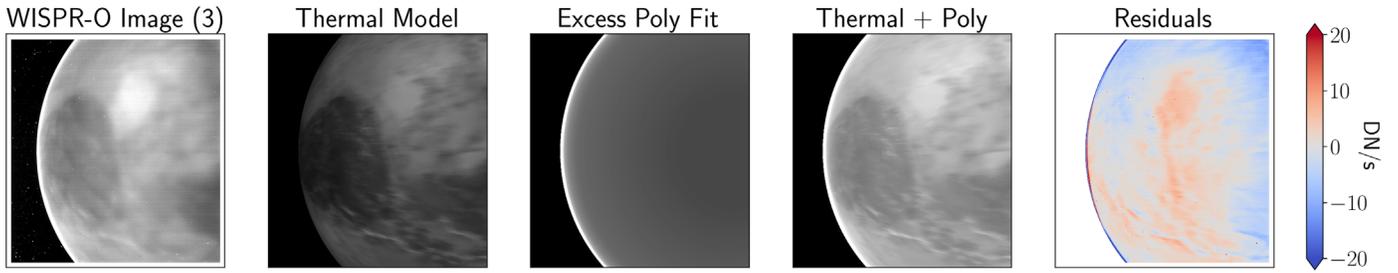

**Figure 10.** WISPR-O image model of Venus. The first panel shows the third WISPR image in the flyby sequence. The second panel shows our thermal emission models projected onto the WISPR image and contains the characteristic surface features seen in the images. The third panel shows a polynomial fit in emission angle to the residuals of the first two panels and contains the limb brightening component. The fourth panel shows the sum of the second and third panels and represents our model of the full WISPR-O image. The fifth panel shows the residuals between the WISPR-O image (first panel) and our image model (fourth panel) with red (blue) indicating an observed brightness excess (deficit) relative to the model. The residuals are Gaussian distributed as $-0.3 \pm 3.5$ DN s$^{-1}$.

To account for the missing nightglow component at the limb, we fit a simple model to the residuals between the WISPR-O image and the thermal model. We take the residuals as a function of emission angle and smooth them using a rolling median with a window of size 150 points. The smoothed residuals exhibit a slight linear increase with emission angle from $0°$ to $65°$, before sharply increasing $>65°$. We use a piecewise analytic model composed of a linear portion and a polynomial portion to fit for the average nightglow contribution. The functional form of the nightglow model is

$$y_{\text{glow}} = \begin{cases} m(\theta - \theta_0) + b, & \text{for } \theta \leqslant \theta_0 \\ \sum_{k=0}^{N} a_k \mu^k, & \text{for } \theta > \theta_0 \end{cases} \quad (2)$$

where $\theta$ is the emission angle, $\mu$ is the cosine of the emission angle, $\theta_0$ is the angle of the break from linear to polynomial, and $m$ (the linear slope) and the set of polynomial coefficients $a_k$ are all fitting parameters. The y-intercept $b$ is determined by the value of the polynomial function at $\theta_0$. We determined that $\theta_0 = 67°$, and $N = 7$ best captures the average residual trend without overfitting. The relatively high polynomial order is required to fit the sharp rise in brightness. Table 1 lists the best fitting model parameters determined using a nonlinear least squares fit to the smoothed residuals using the Python code `scipy.optimize.curve_fit`.

The third panel in Figure 10 shows the residual fit to the median excess emission projected back onto the image plane. The fitted nightglow emission model is able to capture the baseline flux contributions near the center of the Venus disk at low emission angles (as discussed in Section 3.3) and, critically, the stark brightening seen at the Venus limb.

The fourth panel in Figure 10 displays our final WISPR-O image model with the thermal only model from the second panel added to the fitted model to the mean excess brightness from nightglow. This model demonstrates that the WISPR-O image is well explained by emission from the hot Venus surface escaping through the new atmospheric window in the optical, with an overlying emission component from the atmosphere that is dominant at high emission angles, but present across the entire disk.

**Table 1**
Nightglow Model Parameters

| Model Parameter | Value |
| --- | --- |
| $m$ | $1.03 \times 10^{-1}$ |
| $a_0$ | $-8.14 \times 10^6$ |
| $a_1$ | $1.56 \times 10^7$ |
| $a_2$ | $-1.26 \times 10^7$ |
| $a_3$ | $5.52 \times 10^6$ |
| $a_4$ | $-1.42 \times 10^6$ |
| $a_5$ | $2.14 \times 10^5$ |
| $a_6$ | $-1.77 \times 10^4$ |
| $a_7$ | $6.37 \times 10^2$ |

The fifth and final panel of Figure 10 shows the residuals calculated as the WISPR-O image from the first panel minus the final image model in the fourth panel. A diverging colormap is used to highlight regions of Venus where the WISPR data are brighter than the model (red) and darker than the model (blue). Overall, however, the residuals are Gaussian distributed around zero with a standard deviation of around 3.5 DN s$^{-1}$ at $1\sigma$ ($-0.3 \pm 3.5$ DN s$^{-1}$). This indicates that the fit is quite good at capturing the average behavior of the images.

The regions of Venus with $>2\sigma$ deviations from the model are interesting to note and may point to a variety of different factors. First, the brightest surface region in the WISPR-O image—a low elevation lava plain—is also the brightest region of the surface thermal model, but the model is unable to reach the brightness levels seen in the data. This could be due to surface compositional information, cloud optical depth variations, or nightglow spatial structure, which we discuss in Section 4.2. Second, there are *ridges* in the lower third of Venus that are not well fit, and this could result from atmospheric or motion blurring, or small errors in the model used to project the Venus reference data sets onto the WISPR-O images. For particularly small surface features, small shifts in the relative position of the WISPR images and reference data can cause excess residuals. Third, the limb of Venus exhibits some of the largest residuals. This could be explained by a combination of the aforementioned image-model offsets, but the limb residuals also show flux excesses (red) and flux deficits (blue) indicating that brightness variations along the limb that cannot be captured by our average limb brightening model also contribute to the residuals. Such brightness variations along the limb may indicate a spatial structure in the nightglow emission, cloud opacity, scattered sunlight, or a complex combination of these factors. Finally, the rightmost part of the image, particularly the upper right, shows a brightness deficit compared to the models that could be caused by vignetting imperfections not fully removed by the data calibration procedure.

## 4. Discussion

### 4.1. Implication for Surface Geology

WISPR-O measurements of geologic units at similar elevations show distinct differences in brightness that can be attributed to factors that are independent of temperature and emission angle (Figure 7). The Lava unit has the highest brightness values of all mapped units, including the Plains unit (30% at $4\sigma$) and Haasttse-baad tessera (25% at $2.9\sigma$). Although less statistically significant, Ovda tessera is notably brighter than Thetis tessera by 20% at $1.6\sigma$ where both units cover the same elevations and emission angles. These observations demonstrate that the WISPR-O data may be sensitive to compositional or grain size variations on the Venus surface.

Laboratory work shows a positive correlation between 1 $\mu$m emissivity and FeO (ferrous iron) content in rocks at Venus temperatures (440°C; Dyar et al. 2020; Helbert et al. 2021). These measurements are made at longer wavelengths (0.86–1.18 $\mu$m) than the WISPR-O broadband spectral range (0.48–0.80 $\mu$m; Wood et al. 2022). However, Helbert et al. (2021) demonstrate that the high temperature laboratory emissivity measurements of basalts correlate with emissivity derived from photometer data of the Venus surface collected by the Venera 9 and 10 landers at 5 channels over the range 0.5–1.1 $\mu$m (Ekonomov et al. 1980). This implies that higher values of WISPR-O brightness should correspond to greater FeO content in observed rocks. The weathering processes in the deep atmosphere of Venus are modeled to convert ferrous iron to ferric iron on geological timescales of days to tens of thousands of years (e.g., Smrekar et al. 2010; Berger et al. 2019; Filiberto et al. 2020; Radoman-Shaw et al. 2022; Santos et al. 2023). If so, the high WISPR-O brightness of the Lava unit shows that it has the greatest FeO values of the studied units. The Lava unit having greater FeO values than the Plains unit, which are also interpreted to be basaltic lava flows (Hansen 2009), indicates that the Lava unit has experienced less weathering and is thus younger than the Plains flows, or that the Lava flows has an intrinsically higher FeO content. Geomorphologically, the Sogolon Planitia region that corresponds to the Lava unit has smooth, homogeneous volcanic flow materials with a lower density of small volcanic shield features than the adjacent shield terrain to the north. Niobe Planitia to the northeast corresponds to the Plains unit, and is consistent with a different mode and/or timing of emplacement. The high WISPR-O brightness of the smooth flow region may suggest that the Lava unit has a higher FeO content than surrounding shield plains, as well as the Haasttse-baad tessera (see Figure 7). Northern Sogolon Planitia lies partially beneath the parabolic ejecta of the impact crater Merit Ptah. Parabolic ejecta deposits are ephemeral, with a mean crater retention age of ca. 10s Ma (Phillips et al. 1991; Izenberg et al. 1994; Campbell et al. 2007). It is possible that Merit Ptah impact ejecta is geologically young and thus has suffered less weathering than surrounding plains units resulting in a slightly higher FeO content and corresponding WISPR-O brightness.

Thetis and Ovda tesserae have distinct WISPR-O brightness values indicating that these regions, of unknown rock type, have different compositions. The lower brightness values for Thetis indicate a lower FeO content than those of Ovda Regio. This may be due to intrinsic differences in the rocks themselves, where Thetis has a more silica-rich composition than does Ovda, or it may suggest that ferrous iron in Thetis has been preferentially consumed due to differences in the style, rate, or duration of surface-atmosphere weathering of these rocks as compared to Ovda (e.g., via the production of ferric hematite as has been suggested to explain spectral measurements collected by the Venera 9 and 10 landers; Pieters 1983), or enhanced distribution of basaltic crater ejecta on the surface of Ovda as compared to Thetis (Whitten & Campbell 2016). The colinearity of the Thetis–Haasttse-baad WISPR-O versus elevation trends may imply that the two tessera regions are more similar in composition than either are to Ovda. Variations in the radiophysical properties of tesserae also show that tessera composition is not uniform across the planet (Whitten & Campbell 2016; Brossier & Gilmore 2020); the WISPR-O data provide an independent method to assess this variability and its causes.

Particle size has a demonstrated effect on NIR reflectance, where finer grain sizes at the 10s–100s $\mu$m scale will increase reflectance and therefore lower emissivity (e.g., Pieters 1983). If grain size is the dominant cause of WISPR-O brightness, this would suggest that the materials of the surface of the Lava unit have a larger average grain size in the uppermost 10s of $\mu$m than both the Plains and Haasttse-baad units and that Ovda Regio has a larger average grain size than Thetis Regio. Grain sizes are reduced by chemical and physical weathering and/or the addition of sediment, such as from impact ejecta. If due to weathering, this would imply that the Plains and Haasttse-baad and/or Thetis have undergone more extensive weathering due to friability, age, and/or topography than the Lava and/or Ovda units. However, we note that limited measurement of the 1 $\mu$m emissivity of powders and slabs at Venus temperatures to date shows no systematic dependence on particle size, concluding that FeO content is the dominant contributor to emissivity (Helbert et al. 2021).

### 4.2. Caveats and Remaining Uncertainties

A few factors limit the alignment precision between our reference Venus maps and the observed WISPR images, including blurring from nonnegligible spacecraft motion and atmospheric scattering, and uncertainty in the pointing and/or image projections. Although we introduced a blur in the direction of spacecraft motion based on the duration of the exposures and we were conservative in selecting the interiors of known geological units, residual errors may remain and are difficult to quantify. While we did account for motion blur, we did not correct for the atmospheric scattering footprint, and as a result, our image model appears more sharp than the real images. Additionally, trends with surface quantities are likely broadened by this additional and unavoidable scattering uncertainty. Since shorter wavelengths experience greater levels of Rayleigh scattering (Knicely & Herrick 2020), we would expect these optical measurements to have a scattering footprint around or above the nominal 50–100 km footprint in the NIR measured from orbit (Moroz 2002). However, our finding that the cloud opacity in the WISPR band is distinctly lower than the cloud opacity in the 1 μm window could lead to slight improvements in the spatial resolution, but Rayleigh scattering will still be a limiting factor.

We also identified a characteristic of the image projections wherein the alignment between the Venus reference data projections onto the WISPR images were seen to become progressively more misaligned as the flyby advanced. This could be caused by minor pointing errors, SPICE kernel errors, or errors in the FOV projection. No corrections were applied to the observed images or in our analyses to account for alignment artifacts because this effect was insignificant in the WISPR-O images that had Venus centered in the frame, and primarily affected the frames late in the flyby with only the Venus limb visible. This issue also affected the WISPR-I images, and to a larger degree than WISPR-O. Although we were able to reproduce many of our results with the WISPR-I images, the projection alignment issue was severe enough in WISPR-I to warrant further investigation that is beyond the scope of this paper.

There remain degeneracies between whether nightglow or clouds are responsible for the excess flux required across the Venus disk relative to our thermal emission models. For example, we attributed the excess flux seen at low emission angles and high elevations in Figure 9 to nightglow at disk center that rises sharply toward the limb (e.g., in Figure 10). However, the bright limb may be a red herring. The excess flux near disk center could also be, in part, due to suboptimal cloud opacity in the thermal emission models. Specifically, if the optical depth of the $H_2SO_4$ clouds is substantially decreased relative to our nominal cloud model, then the net result would be stronger emission from the surface seen in the WISPR band (see Appendix). However, our radiative transfer model is sufficiently insensitive to small changes in cloud opacity. If the flux offset were entirely attributed to optically thinner clouds, the entire cloud model would need to have roughly half the nominal optical depth to provide a sufficient brightness enhancement to match the WISPR measurements. Moreover, these relative opacity changes linearly impact the thermal emission from the surface, which provides a poor fit to the observed WISPR-O versus surface elevation trend, whereas a constant flux offset that is more indicative of an emission source provides an excellent fit to the measurements (Figure 8). Therefore, the additional flux component near the Venus disk center ($\sim$10 DN s$^{-1}$) is unlikely to be entirely attributable to thinner-than-expected clouds.

Nightglow is clearly needed to capture the limb brightening, and is a plausible explanation for the excess disk brightness. Wood et al. (2022) predicted the WISPR brightness of $O_2$ nightglow at 0.76 μm should be around 1.6 DN s$^{-1}$ based on the typical limb-to-disk brightness ratio of the 1.27 μm feature (Gérard et al. 2008). Our derived brightness excess of $\sim$10 DN s$^{-1}$ well exceeds this estimate. This could indicate a few different effects such as (1) higher intensity disk emission relative to the limb for the 0.76 μm line, or (2) a combination of multiple in-band nightglow lines, for example, from both $O_2$ and the atomic O I 5577 Å green line, or (3) unidentified systematic effects. Observations of oxygen nightglow show substantial spatial and temporal variability (Allen et al. 1992; Crisp et al. 1996; Hueso et al. 2008). This variability could help to explain spatial residuals seen in Figure 10, but the residuals remain relatively small—about 35% intensity variations on a baseline effect of order 10 DN s$^{-1}$ (at 1$\sigma$)—compared to previously observed nightglow heterogeneity with up to 10× contrasts across the nightside (Gérard et al. 2008). A perfect explanation remains elusive. Ultimately, the reliance on only a single photometric band in this work provides limited evidence with which to break the degeneracies with nightglow. Additional observations, particularly concurrent spectroscopy or narrowband imaging, as well as further analyses would be helpful in the future.

### 4.3. Future Flybys and Observation Opportunities

PSP will perform its final flyby of the Venus nightside on 2024 November 6 (VGA7). During this encounter, PSP will pass at $\sim$340 km of the Venus surface at the closest approach, and the event will last approximately 8.5 minutes. VGA7 is expected to probe in the general vicinity of Phoebe Regio, a completely different area of the Venusian surface compared to the VGA4 images studied in this paper. While there may be challenges in directly comparing VGA4 (this work) with VGA7 due to the differing planetary and spacecraft environments, relative differences may still be quite informative. For example, the different surface coverage under VGA7, and the flyby's lower closest approach may allow a comparison of new territory—some quite uniform—over a range of emission angles, which, when compared to Magellan radar data, may aid in deconvolving variable atmospheric effects from the WISPR images. These upcoming WISPR observations will provide one final opportunity to access the Venus surface through the unique *WISPR window* and to further test and validate the findings presented here in advance of dedicated forthcoming Venus missions.

Additionally, upcoming Venus missions will have an incredible opportunity to advance the state of knowledge of the Venus surface. The regions imaged by WISPR-O were not included in the NIR survey of the southern hemisphere provided by Venus Express; thus, the variations in the NIR properties of geologic units imaged by WISPR-O indicate that the global multiband mapping to be provided by the DAVINCI, VERITAS, and EnVision missions will critically advance our understanding of the diversity, origin and relative age of geologic units on Venus.

### 5. Conclusions

The WISPR brightness images are well explained by emission from the hot Venus surface escaping through a new atmospheric window in the optical, with minimal spatial variations due to cloud heterogeneity, and an overlying component of emission from the atmosphere that is most consistent with $O_2$ nightglow. The surface thermal emission correlates strongly with surface elevation (temperature) and emission angle, and weakly with the thermal emissivity of the surface. While strongest at the limb, the nightglow may persist across the entire nightside disk.

WISPR observations of the nightside of Venus present a new tool for the study of Venus' surface potentially linked to compositional distinctions between geologic units. The thermal emissivity correlations may be the key to identifying distinct surface materials, both in terms of age and signatures of weathering, and of initial composition (e.g., FeO content). The WISPR-O images of the 2021 VGA flyby indicate that tessera terrains in Ovda Regio and Thetis Regio may be compositionally distinct, with Ovda having a higher iron content than that of Thetis. Haasttse-baad Tessera appears more compositionally similar to Thetis than to Ovda. In the lower elevations, the smooth Lava unit of Sogolon Planitia has a higher FeO content and thus is potentially less weathered and younger than the surrounding shield terrains of Niobe Planitia. These data confirm that the WISPR observations shortward of ∼0.8 $\mu$m are sensitive to surface characteristics. They also presage the potential compositional diversity of the terrains of Venus that will be revealed by global NIR observations collected by the three upcoming missions to the planet.

## Acknowledgments

We acknowledge internal research and development funding from the Johns Hopkins Applied Physics Laboratory. A.V. and R.A.H. were supported by WISPR Phase-E funding. P.H. was supported by the NASA Parker Solar Probe Program Office for the WISPR program (contract NNG11EK11I) and the Office of Naval Research. Parker Solar Probe was designed, built, and is now operated by the Johns Hopkins Applied Physics Laboratory as part of NASA's Living with a Star (LWS) program (contract NNN06AA01C). Support from the LWS management and technical team has played a critical role in the success of the Parker Solar Probe mission. The WISPR instrument was designed, built, and is now operated by the US Naval Research Laboratory in collaboration with Johns Hopkins University/Applied Physics Laboratory, California Institute of Technology/Jet Propulsion Laboratory, University of Goettingen, Germany, Centre Spatiale de Liege, Belgium, and University of Toulouse/Research Institute in Astrophysics and Planetology. We thank two anonymous reviewers for their thoughtful review of our manuscript.

*Facilities*: NASA's Parker Solar Probe (PSP).

*Software*: Astropy (Robitaille et al. 2013), SpicePy (Annex et al. 2020), numpy (Travis 2006), skimage (van der Walt et al. 2014), matplotlib (Hunter 2007), PySynphot (Lim et al. 2015), pandas (McKinney 2010), imageio (Silvester et al. 2020), Astroquery (Ginsburg et al. 2019), SMART (Meadows & Crisp 1996; Crisp 1997), LBLABC (Meadows & Crisp 1996; Crisp 1997), corner (Foreman-Mackey 2016).

## Data Availability

The PSP/WISPR images studied here are located in the following WISPR repository https://wispr.nrl.navy.mil/data/rel/fits/L2/highcadence/20210220/. Images defining the geologic units described in this study can be found at Lustig-Yaeger et al. (2023). Magellan Global data sets are available at the USGS repository https://astrogeology.usgs.gov/search?pmi-target=venus and are described in Ford et al. (1993).

## Appendix
## Sensitivity to Cloud Opacity Variations

Although cloud optical depth variations were not included in our full radiative transfer grid (see Section 2.4), we briefly explored their impact on the WISPR-O measurements. We ran a series of spectral models across a grid in optical depth scale factors from 0× the nominal optical depth profile (no opacity) to 2× the nominal. This test was performed for each particle mode constituting our Venus cloud model, including Mode 1, Mode 2, Mode 2', Mode 3, the unknown UV absorber, and all particles at once (Crisp 1986; Arney et al. 2014). We convolved all resulting thermal spectra with the WISPR bandpass to calculate the relative change to the WISPR brightness with relative changes to the nominal cloud optical depth.

**Table 2**
Sensitivity to Relative Changes in Cloud Opacity

| Model Parameter | WISPR-O Relative at $\tau = 0$ | 1.0 $\mu$m Relative at $\tau = 0$ | WISPR-O/1.0 $\mu$m |
| --- | --- | --- | --- |
| Mode 1 | 1.04× | 1.04× | 0.0% |
| Mode 2 | 1.18× | 1.26× | 6.3% |
| Mode 2' | 1.12× | 1.19× | 5.9% |
| Mode 3 | 1.14× | 1.20× | 5.0% |
| UV absorber | 1.20× | 1.21× | 0.8% |
| All together | 2.42× | 3.75× | 35.5% |

Table 2 summarizes the results of our cloud sensitivity tests for WISPR-O in comparison to the 1 $\mu$m opacity window. Increases in brightness are shown relative to the nominal cloud model when the opacity is set to zero ($\tau = 0$). We find that each particle mode exhibits linear changes in relative intensity with the opacity scaling factor, independent of observer zenith angle (Table 2 shows results only at 21°.5) and surface temperature.

In general, Table 2 shows that the clouds are more transparent at shorter wavelengths, and therefore, opacity changes are less significant in the WISPR band compared to the 1 $\mu$m window (by about 35%). If all clouds were removed from the model, the brightness of the surface thermal emission in the WISPR bandpass would only increase by a factor of 2.4×. The more subtle and realistic changes to the cloud opacity due to spatial variations are unlikely to cause significant spatial variations in the WISPR images. The proportionally large effect of the unknown UV absorber is potentially misleading, as it results from the modeling assumption that it behaves the same as the submicron Mode 1 particles at wavelengths beyond the blue optical. These findings reflect both the wavelength-dependent scattering properties of each particle mode (see Arney et al. 2014, Figure 2) and their respective nominal optical depth profiles (see Crisp 1986).

Similar calculations to those presented here have been performed in past studies. Moroz (2002) simulated the Venus atmospheric visibility down to 0.65 $\mu$m, but these shortwave calculations were performed only on the illuminated dayside. Where Moroz (2002) briefly discussed cloud opacity variations, it is in the context of differences between the Venera 13 and 14 landing sites. Their results show the smallest relative changes in upwelling flux at the shortest wavelengths, which is in general agreement with our findings. The enhanced Rayleigh scattering optical depth at seen at shorter wavelengths in the Venus atmosphere in the Moroz (2002) models is also present in our models, but is divided out in the relative comparison of different cloud opacity models (which exhibit the same Rayleigh optical depth) in Table 2. Therefore, our result of optically thinner Venus cloud models at shorter wavelengths stands independent of the Rayleigh scattering optical depth trends in the optical. Furthermore, Hashimoto & Sugita (2003) discussed the critical role of multiple reflections between the Venus atmosphere and surface for nightside thermal emission observations, which also likely impacts the WISPR observations. While these effects are considered locally in our radiative transfer model (Meadows & Crisp 1996; Crisp 1997), surface compositional variability is not accounted for in these secondary reflections off the clouds (i.e., the ambient illumination environment reflects a singular surface composition). However, the resultant net reduction in spatial contrast in nightside images discussed in Hashimoto & Sugita (2003) may be a less significant factor at the shorter wavelengths observed by WISPR due to the diminished opacity from the clouds presented and discussed here. This may be contributing to WISPR's stark sensitivity to the Venus surface and the observed spatial contrasts in the images.

## ORCID iDs

J. Lustig-Yaeger https://orcid.org/0000-0002-0746-1980
N. R. Izenberg https://orcid.org/0000-0003-1629-6478
M. S. Gilmore https://orcid.org/0000-0001-8583-1513
L. C. Mayorga https://orcid.org/0000-0002-4321-4581
E. M. May https://orcid.org/0000-0002-2739-1465
A. Vourlidas https://orcid.org/0000-0002-8164-5948
P. Hess https://orcid.org/0000-0003-1377-6353
B. E. Wood https://orcid.org/0000-0002-4998-0893
R. A. Howard https://orcid.org/0000-0001-9027-8249
N. E. Raouafi https://orcid.org/0000-0003-2409-3742
G. N. Arney https://orcid.org/0000-0001-6285-267X

## References

Acton, C., Bachman, N., Semenov, B., & Wright, E. 2018, P&SS, 150, 9
Acton, C. H. 1996, P&SS, 44, 65
Allen, D. A., & Crawford, J. W. 1984, Natur, 307, 222
Allen, D., Crisp, D., & Meadows, V. 1992, Natur, 359, 516
Annex, A., Pearson, B., Seignovert, B., et al. 2020, JOSS, 5, 2050
Arney, G., Meadows, V., Crisp, D., et al. 2014, JGRE, 119, 1860
Baines, K. H., Bellucci, G., Bibring, J.-P., et al. 2000, Icar, 148, 307
Basilevsky, A. T., Shalygin, E. V., Titov, D. V., et al. 2012, Icar, 217, 434
Berger, G., Cathala, A., Fabre, S., et al. 2019, Icar, 329, 8
Brossier, J., & Gilmore, M. S. 2020, Icar, 355, 114161
Campbell, D. B., Stacy, N. J. S., Newman, W. I., et al. 2007, JGR, 97, 16249
Carlson, R. W., Baines, K. H., Encrenaz, T., et al. 1991, Sci, 253, 1541
Crisp, D. 1986, Icar, 67, 484
Crisp, D. 1997, GeoRL, 24, 571
Crisp, D., Meadows, V. S., Bézard, B., et al. 1996, JGR, 101, 4577
Crisp, D., Sinton, W. M., Hodapp, K. W., et al. 1989, Sci, 246, 506
Drossart, P., Bézard, B., Encrenaz, T., et al. 1993, P&SS, 41, 495
Dyar, M. D., Helbert, J., Maturilli, A., Müller, N. T., & Kappel, D. 2020, GeoRL, 47, e90497
Ekonomov, A., Golovin, Y., & Moshkin, B. 1980, Icar, 41, 65
Filiberto, J., Trang, D., Treiman, A. H., & Gilmore, M. S. 2020, SciA, 6, eaax7445
Ford, J. P., Plaut, J. J., Weitz, C. M., et al. 1993, Guide to Magellan Image Interpretation, JPL Publication 93-24 (Pasadena, CA: NASA/JPL), https://history.nasa.gov/JPL-93-24/jpl_93-24.htm
Ford, P. G., & Pettengill, G. H. 1992, JGR, 97, 13103
Foreman-Mackey, D. 2016, JOSS, 1, 24
Fox, N. J., Velli, M. C., Bale, S. D., et al. 2016, SSRv, 204, 7
Gérard, J. C., Saglam, A., Piccioni, G., et al. 2008, GeoRL, 35, L02207
Gilmore, M. S., Mueller, N., & Helbert, J. 2015, Icar, 254, 350
Ginsburg, A., Sipőcz, B. M., Brasseur, C. E., et al. 2019, AJ, 157, 98
Gordon, I. E., Rothman, L. S., Hill, C., et al. 2017, JQSRT, 203, 3
Hansen, V. L. 2009, Geologic Map of the Niobe Planitia Quadrangle (V-23), Venus Scientific Investigations Map 3025, USGS
Hashimoto, G. L., Roos-Serote, M. C., Sugita, S., et al. 2008, JGRE, 113, E00B24
Hashimoto, G. L., & Sugita, S. 2003, JGRE, 108, 5109
Helbert, J., Maturilli, A., Dyar, M. D., & Alemanno, G. 2021, SciA, 7, eaba9428
Hess, P., Howard, R. A., Stenborg, G., et al. 2021, SoPh, 296, 94
Hueso, R., Sánchez-Lavega, A., Piccioni, G., et al. 2008, JGRE, 113, E00B02
Hunter, J. D. 2007, CSE, 9, 90
Husser, T. O., Wende-von Berg, S., Dreizler, S., et al. 2013, A&A, 553, A6
Iwagami, N., Sakanoi, T., Hashimoto, G. L., et al. 2018, EP&S, 70, 6
Izenberg, N. R., Arvidson, R. E., & Phillips, R. J. 1994, GeoRL, 21, 289
Klose, K. B., Wood, J. A., & Hashimoto, A. 1992, JGR, 97, 16353
Knicely, J., & Herrick, R. R. 2020, P&SS, 181, 104787
Lim, P. L., Diaz, R. I., & Laidler, V. 2015, PySynphot User's Guide, Astrophysics Source Code Library (Baltimore, MD: STScI), https://pysynphot.readthedocs.io/en/latest/
Lorenz, R. D., Crisp, D., & Huber, L. 2018, Icar, 305, 277
Lustig-Yaeger, J., Izenberg, N. R., Gilmore, M. S., et al. 2023, Location of "Lava', "Plains", "Haastte-baad", "Ovda", and "Thetis" Mapped Units on the Surface of Venus That Are Presented in Lustig-Yaeger et al. "A WISPR of the Venus Surface: Analysis of the Venus Nightside Thermal Emission at Optical Wavelengths", doi:10.25438/wes02.22867403
McKinney, W. 2010, in Proc. of the 9th Python Sci. Conf., ed. S. van der Walt & J. Millman, 56
Meadows, V. S., & Crisp, D. 1996, JGR, 101, 4595
Moroz, V. I. 2002, P&SS, 50, 287
Moroz, V. I., & Zasova, L. V. 1997, AdSpR, 19, 1191
Mueller, N., Helbert, J., Hashimoto, G. L., et al. 2008, JGRE, 113, E00B17
Mueller, N. T., Helbert, J., Erard, S., Piccioni, G., & Drossart, P. 2012, Icar, 217, 474
Peralta, J., Lee, Y. J., McGouldrick, K., et al. 2017, Icar, 288, 235
Perryman, M. A. C., Lindegren, L., Kovalevsky, J., et al. 1997, A&A, 323, L49
Phillips, R., Schaber, G., Arvidson, R., et al. 1991, Sci, 252, 288
Pieters, C. M. 1983, JGR, 88, 9534
Pollack, J. B., Dalton, J. B., Grinspoon, D., et al. 1993, Icar, 103, 1
Radoman-Shaw, B. G., Harvey, R. P., Costa, G., et al. 2022, M&PS, 57, 1796
Raouafi, N. E., Matteini, L., Squire, J., et al. 2023, SSRv, 219, 8
Robitaille, T. P., Tollerud, E. J., Greenfield, P., et al. 2013, A&A, 558, A33
Santos, A. R., Gilmore, M. S., Greenwood, J. P., et al. 2023, JGRE, 128, e2022JE007423
Seiff, A. 1987, AdSpR, 7, 323
Shalygin, E. V., Markiewicz, W. J., Basilevsky, A. T., et al. 2015, GeoRL, 42, 4762
Silvester, S., Tanbakuchi, A., Müller, P., et al. 2020, imageio/imageio v2.8.0, Zenodo, doi:10.5281/zenodo.3674133
Smrekar, S. E., Stofan, E. R., Mueller, N., et al. 2010, Sci, 328, 605
Stenborg, G., Howard, R. A., Hess, P., & Gallagher, B. 2021, A&A, 650, A28
Thompson, W. T., & Wei, K. 2010, SoPh, 261, 215
Travis, E. 2006, A Guide to NumPy (USA: Trelgol Publishing)
van der Walt, S., Schönberger, J. L., Nunez-Iglesias, J., et al. 2014, PeerJ, 2, e453
Vourlidas, A., Howard, R. A., Plunkett, S. P., et al. 2016, SSRv, 204, 83
Wenger, M., Ochsenbein, F., Egret, D., et al. 2000, A&AS, 143, 9
Whitten, J. L., & Campbell, B. A. 2016, Geo, 44, 519
Wood, B. E., Hess, P., Lustig-Yaeger, J., et al. 2022, GeoRL, 49, e96302